\documentclass[twocolumn,showpacs,preprintnumbers,amsmath,amssymb,superscriptaddress]{revtex4}
\usepackage[maccyr]{inputenc}
\usepackage[T2A]{fontenc}
\usepackage[english, russian]{babel}
\usepackage{bm,textcomp}
\usepackage{epsfig,amssymb,amsmath,bm}
\usepackage[dvipsnames]{color}

\begin{document}

\title{An alternative description of the weak lepton interaction}
\author{K. S. Karplyuk}
\email{karpks@hotmail.com}
 \affiliation{Department of Radiophysics, Taras Shevchenko University, Academic
Glushkov prospect 2, building 5, Kyiv 03122, Ukraine}
\author{O. O. Zhmudskyy}\email{ozhmudsky@physics.ucf.edu}
 \affiliation{Department of Physics, University of Central Florida, 4000 Central Florida Blvd. Orlando, FL, 32816 Phone: (407)-823-4192}
\begin{abstract}
The description of the massive vector- and pseudo-vector bosons is proposed, which allows the use of them as interaction carriers.
The boson propagators guarantee the renormalizability of the theory with massive intermediate bosons. The proposed description is free from
the Englert-Brout-Higgs mechanism. These bosons are used  to describe the weak lepton interactions. In such a 
description boson's and lepton's interaction  Lagrangian coincides with the boson's   
and lepton's interaction Lagrangian in the Glashow-Weinberg-Salam model, which is in good
agreement with the experimental results.
\end{abstract}


\maketitle
\section{Introduction}
The search for a description of the weak interaction usually begins with  the two statements \cite{gr,ch,h,com,o}.
Firstly, the weak interaction is transmitted by means of the massive vector and pseudo-vector bosons, because this interaction 
is of a very short range.  Secondly, the theory with massive intermediate bosons is not renormalizable.
The reconciliation between the first statement and the second statement can be reached with the help of the gauge theory. In this theory,  
the intermediate gauge bosons are initially introduced as massless. Then the gauge bosons acquire a mass by means of the 
Englert-Brout-Higgs mechanism \cite{e,hi,hig}.  It is supposed that they interact
with the invisible Higgs field, which pervades the universe.

It will be shown below that the second statement is not correct. We propose a description of the massive intermediate bosons
which does not break the renormalizability of the theory. With the help of these intermediate bosons we will build a weak lepton
interaction model free of the Englert-Brout-Higgs mechanism and consistent with the experiment.
\section{Intermediate bosons}   
Let us try to describe the weak interaction of the non-conserved pseudo-vector currents via pseudo-vector bosons interchange.
We choose non-conserved currents because the weak interaction currents are non-conserved.   
 
We start from the pseudo-vector non-conserved
currents because Dirac's equation gives us an example of such currents.  From Dirac's equation it follows that pseudo-vector current 
$\bar{\psi}_f\gamma^\mu\gamma^5\psi_i$ does not satisfy the equation of continuity:
\begin{equation}
\partial_\mu(\bar{\psi}_f\gamma^\mu\gamma^5\psi_i)=i2\frac{mc}{\hbar}\bar{\psi}_f\gamma^5\psi_i.
\end{equation}
Here $\gamma^\mu$ --- Dirac's matrix, $\gamma^\mu\gamma^\nu+\gamma^\nu\gamma^\mu=2\eta^{\mu\nu}$, $\gamma^5=i\gamma^0\gamma^1\gamma^2\gamma^3$,
$\psi_f$ and $\psi_i$ --- bispinors, which satisfy Dirac's equation of mass $m$.

Let us start from the massless boson interchange and then proceed to the massive boson interchange.  Although the massless pseudo-vector bosons don't 
{\bf occur} in nature,
the discussion of this option is useful for the massive pseudo-vector bosons consideration.

Imagine that a pseudo-vector current $j^\mu_p$ is conserved. In this case it is possible to describe the massless bosons 
created by using the analogy of the Maxwell equations
\begin{gather}
\partial_\nu G^{\mu\nu}=0,\\
\partial_\nu\tilde{G}^{\mu\nu}=\zeta j^\mu_p,
\end{gather}
{\it in which the pseudo-vector current is used instead of the vector current}. Here $G^{\mu\nu}$ --- the antisymmetric tensor of the second order,
$\tilde{G}^{\mu\nu}$ --- dual to the $G^{\mu\nu}$ tensor
\begin{equation}
\tilde{G}^{\mu\nu}=\frac{1}{2}\varepsilon^{\mu\nu\alpha\beta}G_{\alpha\beta},
\end{equation}
$\varepsilon^{\mu\nu\alpha\beta}$ --- the totally antisymmetric Levi-Civita symbol, $\varepsilon^{0123}=1$, $j^\mu_p=gc\bar{\psi}_f\gamma^\mu\gamma^5\psi_i$,
$g$ --- the hypothetic pseudo-vector currents interaction constant (charge). Equations (2)-(3) are written in the SI system,
constant $\zeta=\sqrt{\frac{\mu_0}{\varepsilon_0}}$ is used for convenience of comparison with the electromagnetic interaction.

We can not use equation (3) if current $j^\mu_p$ is non-conserved for the following reason.  The divergence of the right-hand side 
of the equation (3) is not zero,  $\partial_\alpha j^\alpha_p\neq 0$.  And the divergence of the left-hand side of this equation is 
zero (tensor $G^{\mu\nu}$ is antisymmetric, $\partial_\mu\partial_\nu\tilde{G}^{\mu\nu}\equiv 0$). We can fill this gap using the two 
equations
\begin{gather}
\partial_\nu G^{\mu\nu}=0,\\
\partial^\mu \beta+\partial_\nu\tilde{G}^{\mu\nu}=\zeta j^\mu_p,
\end{gather}
instead of (2)-(3). Here $\beta$ --- the pseudo-scalar field defined by the equation
\begin{equation}
\Box \beta\equiv \partial_\mu\partial^\mu\beta=\zeta\partial_\mu j^\mu_p.
\end{equation}
We can see that field $\beta$ is produced by the non-conserved current ($\partial_\mu j^\mu_p\neq 0$) only, and quantity $j^\mu_p-\partial^\mu\beta/\zeta$
satisfies the continuity equation
\begin{equation}
\partial_\mu \Bigl(j^\mu_p-\frac{1}{\zeta}\partial^\mu\beta\Bigl)=0.
\end{equation}
We can say that  the field gradient current $\partial^\mu\beta/\zeta$, associated with field $\beta$, complements current $j^\mu_p$ so that the total current
$j^\mu_p-\partial^\mu\beta/\zeta$ becomes conserved. Field $\beta$ appears in order to restore the continuity of the total current.  

This situation is similar to electrodynamics.  The displacement current $\frac{1}{c}\frac{\partial\bm{E}}{\partial t}$, which arises at the surface 
of discontinuity of the conduction current (e.g. between the plates of a parallel plate capacitor), recovers the continuity of the total 
current.  

If current  $j^\mu_p$ is conserved ($\partial_\mu j^\mu_p=0$), there is no need to use field  $\beta$.  According to  
equation (7), the field can not be produced by the conserved current. 

Similar to electrodynamics, fields  $G^{\mu\nu}$ and $\beta$ can be expressed via pseudo-vector potential $Z_\alpha$:
\begin{gather}
G^{\mu\nu}=\varepsilon^{\mu\nu\alpha\beta}\partial_\alpha Z_\beta,\hspace{7mm}
\tilde{G}^{\mu\nu}=-(\partial^\mu Z^\nu-\partial^\nu Z^\mu),\\
\beta=\partial_\alpha Z^\alpha.
\end{gather}
In this case, the homogeneous equation (5) is satisfied identically and equation (6) is satisfied if potential $Z^\alpha$ satisfies 
\begin{equation}
\Box Z^\alpha=\zeta j^\alpha_p.
\end{equation}
As we can see, the massless pseudo-vector boson propagator, described by equations (5)-(6) and 
potential $Z^\alpha$, is the same as the photon propagator.
This is why the interaction of non-conserved currents, implemented by these bosons, is renormalizable.

Note that equations (5)-(6) are not invariant under the gauge transformation $Z'_\alpha=Z_\alpha+\partial_\alpha \varphi$. Although the field
$G^{\mu\nu}$ does not change under this transformation, the field $\beta$ does change and gets the additive $\partial_\alpha\partial^\alpha\varphi$.
The field $\beta$ does not change only if $\Box \varphi=0$. There is no reason to require the gauge invariance of the fields, which describe
the interaction of the non-conserved currents, because Noether's theorem connects the gauge invariance with charge conservation.

 Next, we can proceed to the massive bosons. Equations for the massive boson description  were suggested by Proca \cite{pr}. In our case it is
convenient to write them as
\begin{gather}
\partial_\nu\tilde{G}^{\mu\nu}-\varkappa_ZU^\mu=\zeta j^\mu_p,\\
\partial_\nu G^{\mu\nu}=0.
\end{gather}
Here $\varkappa_Z=m_Zc/\hbar$, fields $G^{\mu\nu}$ and $\tilde{G}^{\mu\nu}$ are connected with the potential $Z^\mu$ by (9), and
\begin{equation}
U^\mu=-\varkappa_Z Z^\mu.
\end{equation}
For potential $Z^\mu$ we obtain:  
\begin{equation}
\Box Z_\mu+\varkappa_Z^2Z_\mu=\zeta(\eta_{\mu\nu}-\frac{1}{\varkappa_Z^2}{\partial_\mu\partial_\nu})j_p^\nu.
\end{equation}
The propagator for the corresponding massive boson is
\begin{equation}
D_{\mu\nu}=-\frac{\eta_{\mu\nu}-k_\mu k_\nu/\varkappa_Z^2}{k^2-\varkappa_Z^2}.
\end{equation}
The propagator (16) leads to the theory which is not renormalizable. The cause of this problem is the term $k^\mu k^\nu/\varkappa_Z^2$ in the numerator (16).
It follows from equation (15), that this term arises in the case of non-conserved currents $\partial_\nu j_p^\nu\neq 0$ only. For the conserved currents
($\partial_\nu j_p^\nu= 0$) this term is missing. The cause of these issues is the same as for the massless case 
--- non-conservation of interacting
currents. It does not matter which intermediate bosons we have --- massive or massless. Consequently, in order to eliminate the problem we can use the
same recipe as for the massless case. Namely, we can complement equation (12) by the field $\beta$ and use the following equations
\begin{gather}
\partial^\mu\beta+\partial_\nu\tilde{G}^{\mu\nu}-\varkappa_ZU^\mu=\zeta j^\mu_p,\\
\partial_\nu G^{\mu\nu}=0,\\
\varepsilon^{\mu\nu\alpha\beta}\partial_\alpha U_\beta+\varkappa_ZG^{\mu\nu}=0,\\
\partial_\mu U^\mu+\varkappa_Z\beta=0,
\end{gather}
for the description of the massive pseudo-vector bosons.

The homogeneous equations (18)-(20) are satisfied identically if fields $G^{\mu\nu}$, $\beta$ and $U^\mu$ are connected with potential $Z^\mu$
by equations (9), (10), (14). In order to satisfy equation (17) potential $Z^\mu$ has to be defined by
\begin{equation}
\Box Z^\mu+\varkappa_Z^2Z^\mu=\zeta j_p^\mu.
\end{equation}
The propagator of the corresponding massive  pseudo-vector boson is
\begin{equation}
D^{\mu\nu}=-\frac{\eta^{\mu\nu}}{k^2-\varkappa_Z^2}.
\end{equation}
This form of the propagator provides the renormalizability of the theory, which uses massive  pseudo-vector bosons (17)-(20) as interaction carriers.

Note that we can derive equations (17)-(20) in a usual way.   Variation of the Lagrangian density  
\begin{equation}
\mathcal{L}=\underbrace{\frac{1}{2\zeta c^2}\Bigl(\frac{\tilde{G}_{\alpha\beta}\tilde{G}^{\alpha\beta}}{2}+
\beta^2-U_\alpha U^\alpha\Bigl)}_{\mathcal{L}_Z}+\underbrace{\frac{1}{c^2}j^\alpha_m Z_\alpha}_{\mathcal{L}^Z_{int}},
\end{equation}
by $Z_\alpha$ gives the desired system (17)-(20), where definitions (9), (10), (14) are used. 

\section{Neutral currents interaction}
Let us use bosons, described by equations (17)-(20), as intermediate bosons  for the  interaction of neutral lepton currents. We restrict
ourselves to the discussion of the electrons and electron neutrino interaction because other leptons can be included in an additive way.
Namely, let us discuss the interaction of the pseudo-vector neutral currents:
\begin{gather}
j^\alpha_p=g_Zc(2\bar{\nu}\gamma^\alpha\gamma^5\nu-\bar{e}\gamma^\alpha\gamma^5e).
\end{gather}
Term $2\bar{\nu}\gamma^\alpha\gamma^5\nu$ in (24) can be written in the following way
\begin{gather}
2\bar{\nu}\gamma^\alpha\gamma^5\nu=-\bar{\nu}\gamma^\alpha(1-\gamma^5)\nu+\bar{\nu}\gamma^\alpha(1+\gamma^5)\nu.
\end{gather} 
The first term in (25) describes contribution of the left neutrino in the current $j_p^\alpha$.  The second term describes contribution of the right neutrino.  
Let us assume that neutrino is massless. Only the left neutrino can exist as massless.  That is why we don't need to take into account the 
contribution of the right (not existing) neutrino and term $\bar{\nu}\gamma^\alpha(1+\gamma^5)\nu$ in (25) can be omitted:  
\begin{gather}
j^\alpha_p=g_Zc[-\bar{\nu}\gamma^\alpha(1-\gamma^5)\nu-\bar{e}\gamma^\alpha\gamma^5e].
\end{gather}

Let us use current (26) in the equation (23) and write an expression for Lagrangian density for interacting electrons, left-handed massless neutrinos,
photons and $Z$-bosons as
\begin{gather}
\mathcal{L}=\mathcal{L}_e+\mathcal{L}_\nu+\mathcal{L}_A+\mathcal{L}_Z+\mathcal{L}_{int}^{A}+\mathcal{L}_{int}^{Z}.
\end{gather}
The term $\mathcal{L}_A=-(1/2\zeta c^2)(F_{\alpha\beta}F^{\alpha\beta}/2)$ is the density of electromagnetic Lagrangian, the term $\mathcal{L}_Z$
is expressed by equation (23),
\begin{gather}
\mathcal{L}_e=i\frac{\hbar}{2}(\bar{e}\gamma^\alpha\partial_\alpha e-\partial_\alpha\bar{e}\gamma^\alpha e)-m_ec\bar{e}e,\\
\mathcal{L}_\nu=i\frac{\hbar}{2}\Bigl(\bar{\nu}\gamma^\alpha\partial_\alpha\frac{1-\gamma^5}{2}\nu-\partial_\alpha\bar{\nu}\gamma^\alpha\frac{1-\gamma^5}{2}\nu\Bigr),\\
\mathcal{L}_{int}^{A}=-\frac{1}{c^2}j^\alpha A_\alpha=-\frac{1}{c^2}(q c\bar{e}\gamma^\alpha e)A_\alpha=-\frac{q}{c}\bar{e}\gamma^\alpha e A_\alpha,\\
\mathcal{L}_{int}^{Z}=\frac{1}{c^2}j^\alpha_p Z_\alpha=-\frac{g_Z}{c}\bigl[\bar{\nu}\gamma^\alpha(1-\gamma^5)\nu+\bar{e}\gamma^\alpha\gamma^5e\bigr]Z_\alpha.
\end{gather}
Here $q=-e_0$, $e_0=1,6\cdot 10^{-19}$[C]. We assume that neutrino is massless. The definition of the current (26) corresponds to two extra
terms in Dirac's equations. Term $-\frac{g_Z}{\hbar c}Z_\alpha\gamma^\alpha\gamma^5e$ for the electrons and term
$-2\frac{g_Z}{\hbar c}Z_\alpha\gamma^\alpha\frac{1-\gamma^5}{2}\nu$ for the neutrino (besides the electromagnetic term):
\begin{gather}
i\gamma^\alpha\partial_\alpha e-\frac{q}{\hbar c}A_\alpha\gamma^\alpha e-\frac{g_Z}{\hbar c}Z_\alpha\gamma^\alpha\gamma^5e-\frac{m_e c}{\hbar}e=0,\\
i\gamma^\alpha\frac{1-\gamma^5}{2}\partial_\alpha\nu +2\frac{g_Z}{\hbar c}Z_\alpha\gamma^\alpha\frac{1-\gamma^5}{2}\nu=0.
\end{gather}
The  Lagrangian density of the interaction $\mathcal{L}_{int}^{A}+\mathcal{L}_{int}^{Z}$ nearly coincides with the neutral currents' 
Lagrangian density  of the interaction  in the Glashow-Weinberg-Salam model \cite{g,w,s} if we suppose $g_Z=e_0/4\sin\theta_W\cos\theta_W$:
\begin{gather}
\mathcal{L}_{int}=\mathcal{L}^{int}_{A}+\mathcal{L}^{int}_{Z}-\frac{q}{c}\bar{e}\gamma^\alpha e A_\alpha-\nonumber\\-
\frac{e_0}{4c\sin\theta_W\cos\theta_W}\bigl[\bar{\nu}\gamma^\alpha(1-\gamma^5)\nu +\bar{e}\gamma^\alpha\gamma^5 e\bigr] Z_\alpha.
\end{gather}
The only exception is the term $\frac{e_0}{4c\sin\theta_W\cos\theta_W}(1-4\sin^2\theta_W)\bar{e}\gamma^\alpha e Z_\alpha$ in the 
Glashow-Weinberg-Salam model.  

This term determines the contribution of the vector component of the electron current into the weak interaction of the neutral currents.  
The up-to-date value of the electron current components (in square brackets (34)) is \cite{k}: 
\begin{gather}
-2(0.043\pm 0.063)\bar{e}\gamma^\alpha e+2(0.545\pm 0.056)\bar{e}\gamma^\alpha\gamma^5 e.
\end{gather}  
As we see, experimental data allows to remove the vector component  $\bar{e}\gamma^\alpha e$ of the electron current and to retain the pseudo-vector 
component $\bar{e}\gamma^\alpha\gamma^5 e$ only.  It corresponds to the value $\sin^2\theta_W=0.25$ (i.e. $\theta_W=\pi/6$) in the 
Glashow-Weinberg-Salam model.  In this case, neutral current and Z-bosons become pure pseudo-vectors. The Lagrangian density (34)  and the 
Lagrangian density in the Glashow-Weinberg-Salam model coincide and can be written as:    
\begin{gather}
\mathcal{L}_{int}=\mathcal{L}_{int}^{A}+\mathcal{L}_{int}^{Z}=\nonumber\\=
-\frac{q}{c}\bar{e}\gamma^\alpha e A_\alpha-\frac{e_0}{c\sqrt{3}}\bigl[\bar{\nu}\gamma^\alpha(1-\gamma^5)\nu +\bar{e}\gamma^\alpha\gamma^5 e\bigr] Z_\alpha.
\end{gather}  
Note that for the massive fermions the pseudo-vector bilinear expression $\bar{\psi}\gamma^\alpha\gamma^5\psi$ is in proportion 
to the spin vector $s^\alpha$:
\begin{gather}
\bar{\psi}\gamma^\alpha\gamma^5\psi=s^\alpha|N|^2=\Bigl(\frac{\bm{n}\cdot\bm{p}}{mc}, \bm{n}+\frac{\bm{p}(\bm{n}\cdot\bm{p})}{mc(mc+p_o)}\Bigr)|N|^2.
\end{gather}
Here $|N|^2=\bar{\psi}\psi$ --- normalization factor, $\bm{n}$ --- unit spin vector in fermion's own reference frame.  As we can see, 
the direction of the spin determines the direction of the  neutral current.

The Lagrangian of the Glashow-Weinberg-Salam model is in good agreement with the experimental data. That is why Lagrangian (36)
is ought to be in good agreement with experiments too. Thus, the Lagrangian density (36) can be used for the description of
the neutral currents' interaction as well as the Lagrangian density in Glashow-Weinberg-Salam model. In this case, the boson's mass $m_Z$ and
coupling coefficient $g_Z$ have to be determined from the experiment.

This description is similar to quantum electrodynamics, but instead of massless vector photons the massive pseudo-vector $Z$-bosons
with propagator (22) are used as interaction carriers. The description is renormalizable and does not use the Englert-Brout-Higgs mechanism.

Are the massive Z-bosons purely pseudo-vector particles? The final answer can only be found experimentally. It is also possible that the 
omitted term with multiplier $(1-4\sin^2\theta_W)$ has to be retained and the Z-boson is a vector-pseudovector mixture. 
We will discuss the mixed bosons in the next section (\# IV).  It will be shown in section \#V how we can take into account 
possible vector additives to the Z-boson term.  

\section{Charged currents interaction}
The carriers of the interaction between the charged currents are the charged bosons. The description of the charged bosons is
given by the complex potentials $W_{\pm}^\alpha=X^\alpha\pm iY^\alpha$. Let us first write equations for the fields,
which are connected to the potentials $X^\alpha$ and $Y^\alpha$, and then equations for the complex potentials
$W_{\pm}^\alpha=X^\alpha\pm iY^\alpha$.

The charged currents are non-conserved. Let us use arguments of the previous section and write equations for the fields created by two
non-conserved currents: vector current $j^\mu_v$ and pseudo-vector current $j^\mu_p$
\begin{gather}
\partial^\mu\epsilon-\partial_\nu K^{\mu\nu}-\varkappa_W V^\mu=\zeta j^\mu_v,\\
\partial^\mu\beta+\partial_\nu \tilde{K}^{\mu\nu}-\varkappa_W U^\mu=\zeta j^\mu_p,\\
(\partial^\mu V^\nu-\partial^\nu V^\mu)+\varepsilon^{\mu\nu\alpha\beta}\partial_\alpha U_\beta+\varkappa_W K^{\mu\nu}=0,\\
\partial_\alpha V^\alpha+\varkappa_W\epsilon=0,\\
\partial_\alpha U^\alpha+\varkappa_W\beta=0.
\end{gather} 
These equations are equivalent to the Dirac-Kahler's equations \cite{ka}, if right-hand sides of equations (38)-(39) are zero.  
In three dimensional notation these equations can be written as:
\begin{align}
\frac{1}{c}\frac{\partial {\epsilon}}{\partial t}+\nabla\cdot{\bm E}'-\varkappa_W  V_0&=\zeta j_{v0},\\
\frac{1}{c}\frac{\partial {\bm E}'}{\partial t}-\nabla\times c{\bm B}'+ \nabla\epsilon+\varkappa_W {\bm V}&=-\zeta{\bm j}_v,\\
-\frac{1}{c}\frac{\partial\beta}{\partial t}+\nabla\cdot c{\bm B}'+\varkappa_W  U_0&=-\zeta j_{p0},\\
\frac{1}{c}\frac{\partial c{\bm B}'}{\partial t}+\nabla\times {\bm E}'-\nabla\beta-\varkappa_W {\bm U}&=\zeta{\bm j}_p,\\
\frac{1}{c}\frac{\partial V_0}{\partial t}+\nabla\cdot{\bm V}+\varkappa_W{\epsilon}&=0,\\
\frac{1}{c}\frac{\partial{\bm V}}{\partial t}+\nabla V_0-\nabla\times{\bm U}-\varkappa_W{\bm E}'&=0,\\
\frac{1}{c}\frac{\partial U_0}{\partial t}+\nabla\cdot{\bm U}+\varkappa_W{\beta}&=0,\\
\frac{1}{c}\frac{\partial{\bm U}}{\partial t}+\nabla U_0+\nabla\times{\bm V}+\varkappa_W c{\bm B}'&=0.
\end{align}
Here ${\bm E}'=(-K^{01},-K^{02},-K^{03})$, $c{\bm B}'=(-K^{23},-K^{31},-K^{12})$. Symbol prime in ${\bm E}'$ and $c{\bm B}'$
differs these variables from regular electric and magnetic fields.

Let us connect the fields with vector potential $X^\alpha$ and pseudo-vector potential $Y^\alpha$:
\begin{gather}
K^{\mu\nu}=(\partial^\mu X^\nu-\partial^\nu X^\mu)+\varepsilon^{\mu\nu\alpha\beta}\partial_\alpha Y_\beta,\\
\tilde{K}^{\mu\nu}=\frac{\varepsilon^{\mu\nu\alpha\beta}}{2} K_{\alpha\beta}=\varepsilon^{\mu\nu\alpha\beta}\partial_\alpha X_\beta
-(\partial^\mu Y^\nu-\partial^\nu Y^\mu),\\
V^\mu=-\varkappa_W X^\mu,\hspace{13mm} U^\mu=-\varkappa_W Y^\mu,\\
\epsilon=\partial_\alpha V^\alpha, \hspace{13mm} \beta=\partial_\alpha U^\alpha.
\end{gather}
The homogeneous equations (40)-(42) are satisfied identically. In order to satisfy nonhomogeneous equations (38)-(39) potentials
have to satisfy equations
\begin{gather}
\Box X^\alpha+\varkappa_W^2 X^\alpha=\zeta j^\alpha_v,\\
\Box Y^\alpha+\varkappa_W^2 Y^\alpha=\zeta j_p^\alpha.
\end{gather}
The two potentials $X$ and $Y$ represent one boson with mass $m_W=\hbar\varkappa_W/c$ and propagator
\begin{gather}
D^{\mu\nu}=-\frac{\eta^{\mu\nu}}{k^2-\varkappa_W^2}.
\end{gather}
The propagator guarantees the renormalizability of the theory. Such a two-potential description  of the boson is similar
to the electron description by the Dirac bispinors (combine two spinors into one).

Equations (38)-(42) can be derived by solving the variational problem of Lagrangian density
\begin{gather}
\mathcal{L}=\mathcal{L}_{XY}+\mathcal{L}_{int}^{XY},\\
\mathcal{L}_{XY}=\frac{1}{2\zeta c^2}\Bigl(-\frac{K_{\alpha\beta} K^{\alpha\beta }}{4}+\frac{\tilde{K}_{\alpha\beta} \tilde{K}^{\alpha\beta }}{4}-
\epsilon^2+\beta^2+\nonumber\\+V_\alpha V^\alpha-U_\alpha U^\alpha\Bigr),\\
\mathcal{L}_{int}^{XY}=\frac{1}{c^2}(-j^\alpha_v X_\alpha+j^\alpha_p Y_\alpha),
\end{gather}
with respect to $X_\alpha$ and $Y_\alpha$.

We can separate equations  (38)-(42) into two independent systems and mark them by subscripts \lq\lq-\rq\rq\, and \lq\lq\,+ \rq\rq\,.
Let us introduce variables
\begin{gather}
K^{\pm}_{\alpha\beta}=K_{\alpha\beta}\mp i\tilde{K}_{\alpha\beta},V^{\pm}_{\alpha}= V_{\alpha}\pm iU_{\alpha},
\epsilon^{\pm}=\epsilon\pm i\beta,
\end{gather}
instead of $K_{\alpha\beta}$, $\tilde{K}_{\alpha\beta}$, $V_\alpha$, $U_\alpha$, $\epsilon$, $\beta$. We get
\begin{gather}
\partial_\alpha K^{\alpha\beta}_\pm+\partial^\beta \epsilon_{\pm}-\varkappa_W V^\beta_{\pm}=\zeta(j^\beta_v\pm ij^\beta_p),\\
\partial^\mu V^\nu_\pm-\partial^\nu V^\mu_\pm\mp i\varepsilon^{\mu\nu\alpha\beta}\partial_\alpha V_\beta^\pm+\varkappa_W K^{\mu\nu}_\pm=0,\\
\partial_\alpha V^\alpha_{\pm}+\varkappa_W\epsilon_{\pm}=0.
\end{gather}
Fields with subscripts \lq\lq\, - \rq\rq\,  and \lq\lq\,+ \rq\rq\,  are connected with the potentials $W^{\pm}_\alpha=X_\alpha\pm iY_\alpha$
with the same indices:
\begin{gather}
K^{\mu\nu}_{\pm}=\partial^\mu W^\nu_{\pm}-\partial^\nu W^\mu_{\pm}\mp i\varepsilon^{\mu\nu\alpha\beta}\partial_\alpha W_\beta^{\pm},\\
V^{\pm}_\alpha=-\varkappa_W W^{\pm}_\alpha,\hspace{13mm}\epsilon_{\pm}=\partial_\alpha W_{\pm}^\alpha.
\end{gather}
The potentials $W^{\pm}_\alpha$ are defined from the equations
\begin{gather}
\Box W_{\pm}^\alpha+\varkappa_W^2W_{\pm}^\alpha=\zeta(j^\alpha_v\pm ij^\alpha_p).
\end{gather}

The transition to the scalars, vectors and tensors with indices \lq\lq\, - \rq\rq\,  and  \lq\lq\,+ \rq\rq\,  is similar to the transition to
the left and right spinors in Lorentz's chiral group presentation.  It is the presentation of left and right scalars, vectors and tensors of this group.
In contrast to the left and right spinors, the left and right scalars, vectors and tensors transformations are the same under the proper Lorentz's 
transformation. But under spatial inversion, the left scalars, vectors and tensors turn into the right scalars, vectors and tensors and vice versa.

Let us connect currents $j^\alpha_v$ and $j^\alpha_p$ with electron and neutrino bispinors. We can obtain a better agreement with the experiments 
 in the following way:
\begin{gather}
j^\alpha_v=g_Wc(\bar{\nu}\gamma^\alpha e+\bar{e}\gamma^\alpha \nu),\\
j^\alpha_p=ig_Wc(\bar{\nu}\gamma^\alpha\gamma^5 e-\bar{e}\gamma^\alpha\gamma^5 \nu).
\end{gather}
For the currents $j^\alpha_{+}=j^\alpha_v+ij^\alpha_p$ and $j^\alpha_{-}=j^\alpha_v-ij^\alpha_p$ we obtain
\begin{gather}
j^\alpha_{+}=g_Wc\bigl[\bar{\nu}\gamma^\alpha(1-\gamma^5) e+\bar{e}\gamma^\alpha(1+\gamma^5)\nu)\bigr],\\
j^\alpha_{-}=g_Wc\bigl[\bar{\nu}\gamma^\alpha(1+\gamma^5) e+\bar{e}\gamma^\alpha(1-\gamma^5)\nu)\bigr].
\end{gather}
The terms with the multiplier $(1+\gamma^5)$ describe the contribution of the right neutrino to the currents.  
Because the right-handed massless neutrino does not exist, these terms can be omitted: 
\begin{gather}
j^\alpha_{+}=j^\alpha_v+ij^\alpha_p=g_Wc\bar{\nu}\gamma^\alpha(1-\gamma^5) e,\\
j^\alpha_{-}=j^\alpha_v-ij^\alpha_p=g_Wc\bar{e}\gamma^\alpha(1-\gamma^5)\nu.
\end{gather}
According to (67), potentials $W_{+}^\alpha$ and $W_{-}^\alpha$ are defined from the equations
\begin{gather}
\Box W_{+}^\alpha+\varkappa_W^2W_{+}^\alpha=\zeta g_Wc\,\bar{\nu}\gamma^\alpha(1-\gamma^5) e,\\
\Box W_{-}^\alpha+\varkappa_W^2W_{-}^\alpha=\zeta g_Wc\,\bar{e}\gamma^\alpha(1-\gamma^5)\nu.
\end{gather}
The propagator of the bosons connected with potentials $W_{+}^\alpha$ and $W_{-}^\alpha$ coincides with (57).

Taking into account (72)-(73) Lagrangian density of the charged currents, interaction (60) can be written as
\begin{gather}
\mathcal{L}_{int}^{XY}=\frac{1}{c^2}(-j^\alpha_v X_\alpha+j^\alpha_p Y_\alpha)\equiv\nonumber\\
\equiv\frac{(j^\alpha_v-ij^\alpha_p)(X_\alpha-iY_\alpha)+(j^\alpha_v+ij^\alpha_p)(X_\alpha+iY_\alpha)}{2c^2}=\nonumber\\=
-\frac{g_W}{2c}\bigl[\bar{e}\gamma^\alpha(1-\gamma^5)\nu\,W_\alpha^{-}+\bar{\nu}\gamma^\alpha(1-\gamma^5)e\,W_\alpha^{+}\bigr].
\end{gather}
This Lagrangian density coincides with the Lagrangian density in the Glashow-Weinberg-Salam model if we choose the interaction coefficient
as $g_W=e_0/(\sqrt{2}\sin \theta_W)=e_0\sqrt{2}$. The Lagrangian of the Glashow-Weinberg-Salam model is in good agreement with the 
experiment data. That is why Lagrangian (76) is ought to be in good agreement with the experimental results too.

\section{Discussion}
Let us gather all the results and write the expression for the Lagrangian density of the electron-neutrino interaction via the exchange
of photons, neutral $Z$-bosons and charged $W^{\pm}$-bosons:
\begin{gather}
\mathcal{L}=\mathcal{L}_A+\mathcal{L}_Z+\mathcal{L}_{XY}+\mathcal{L}_e+\mathcal{L}_\nu+\mathcal{L}_{int}.
\end{gather}
Here
\begin{gather}
\mathcal{L}_{int}=\mathcal{L}_{int}^{A}+\mathcal{L}_{int}^{Z}+\mathcal{L}_{int}^{XY}=\nonumber\\=
-\frac{q}{c}\bar{e}\gamma^\alpha e A_\alpha-\frac{g_Z}{c}\bigl[\bar{\nu}\gamma^\alpha(1-\gamma^5)\nu +\bar{e}\gamma^\alpha\gamma^5 e\bigr]Z_\alpha-\nonumber\\
-\frac{g_W}{2c}\bigl[\bar{e}\gamma^\alpha(1-\gamma^5)\nu\,W_\alpha^{-}+\bar{\nu}\gamma^\alpha(1-\gamma^5)e\,W_\alpha^{+}\bigr].
\end{gather}
Here the $Z$- and $W^{\pm}$-boson propagators described by (22) and (57). This guarantees the renormalizability of the theory.

According to the Lagrangian density (77) electron and neutrino are described by equations
\begin{gather}
i\gamma^\alpha\partial_\alpha e-\frac{m_e c}{\hbar}e-\frac{q}{\hbar c}A_\alpha\gamma^\alpha e-\nonumber\\
-\frac{g_Z}{\hbar c}Z_\alpha\gamma^\alpha\gamma^5e
-\frac{g_W}{\hbar c}W_\alpha^{-}\gamma^\alpha\frac{1-\gamma^5}{2}\nu=0,
\end{gather}
\begin{gather}
i\gamma^\alpha\partial_\alpha\frac{1-\gamma^5}{2}\nu-\nonumber\\-2\frac{g_Z}{\hbar c}Z_\alpha\gamma^\alpha\frac{1-\gamma^5}{2}\nu-
\frac{g_W}{\hbar c}W_\alpha^{+}\gamma^\alpha\frac{1-\gamma^5}{2}e=0.
\end{gather}

The Lagrangian interaction density (78) differs from the Lagrangian interaction density in the Glashow-Weinberg-Salam model
by the small term with the vector current $(e_0/4c\sin\theta_W\cos\theta_W)(1-\sin^2\theta_W)\bar{e}\gamma^\alpha eZ_\alpha$.
We neglect this term similar to (31) but it can be easily taken into account. We can do so if we assume that the neutral boson
has not only the pseudo-vector component but the vector component also. The boson is created not only by pseudo-vector current (26) but
also by vector current $j_v^\alpha=g_v c\bar{e}\gamma^\alpha e$. This boson is described by the equations (38)-(42) and by the two potentials:
the pseudo-vector potential $Z^\alpha$ (analogy of $Y^\alpha$) and the vector potential $Z^\alpha_v$ (analogy of $X^\alpha$). The pseudo-vector potential
is created by the pseudo-vector current (24) (equation (56)). The vector potential is created by the vector current
$j_v^\alpha=g_v c\bar{e}\gamma^\alpha e$ (equation (55)). They describe one boson of mass
$m_Z$. This boson propagator is still described by equation (22). The Lagrangian interaction density of the neutral currents becomes
\begin{gather}
\mathcal{L}_{int}^{A}+\mathcal{L}_{int}^{Z}+\mathcal{L}_{int}^{Z_v}=\nonumber\\
=-\frac{q}{c}\bar{e}\gamma^\alpha e A_\alpha-\frac{g_Z}{c}[\bar{\nu}\gamma^\alpha(1-\gamma^5)\nu +\bar{e}\gamma^\alpha\gamma^5 e]Z_\alpha+\nonumber\\+
\frac{g_v}{c}\bar{e}\gamma^\alpha e Z_{v\alpha}.
\end{gather}
Connection coefficient $g_v$ can always be determined in such a way that the calculation results are in good agreement with the experimental data.
In this case, term $\frac{g_v}{c}\bar{e}\gamma^\alpha e Z_{v\alpha}$ correspondents to the term
$(g/4\cos\theta_W)(1-\sin^2\theta_W)\bar{e}\gamma^\alpha  eZ_\alpha$ in the Glashow-Weinberg-Salam model. In any case it is small,  hence 
we neglect it in the equations (31) and (78).

As it was shown, the proposed model can describe any interaction implemented by the interchange of the vector bosons and/or  
by the interchange of the pseudo-vector bosons. Among other things, it can be expanded to the weak interaction of the massive 
neutrino and weak quark interaction.

An agreement between calculation results and experimental results can be achieved by choice of the model parameters:
expressions for the interacted currents, connection coefficients and the boson masses.

The expression of the interacted currents reflects the existence of some regularities in the structure of the interacted
particles' multiplets.

The proposed model does not use the Englert-Brout-Higgs mechanism nor Higgs boson. That is why the question arose about the scalar boson with mass
125\,GeV, that was found not long ago. In the proposed model it is supposed to be one of the scalar bosons. It is not
a problem to include in the model a scalar or pseudo-scalar boson. It can be done easily because system (38)-(42) also describes these
bosons among others.
\section{Acknowledgements}
The authors would like to thank  Prof. Denis Kovalenko for stimulating discussions.

\end{document}